\documentclass[twocolumn,secnumarabic,amssymb, nobibnotes, aps, prl, superscriptaddress]{revtex4-2}
\usepackage{booktabs}
\usepackage{enumitem}
\usepackage{graphicx}% Include figure files
\usepackage{dcolumn}% Align table columns on decimal point
\usepackage{bm}% bold math
\usepackage{amsmath}
\usepackage{lineno}
\usepackage{ragged2e}
\usepackage{enumerate}
\usepackage{setspace}
\usepackage{dsfont}
\usepackage{physics}
\usepackage{subfigure}
\usepackage{multirow}
\usepackage{indentfirst} %段落缩进值宏包
\usepackage {mathrsfs}%花体字
\usepackage{color}%字体颜色
\usepackage[margin=2.3cm]{geometry}
\usepackage{lineno}
\usepackage{amssymb,amsfonts,amsthm,bm,mathtools}
\usepackage{makecell}   %表格
\usepackage{multirow}	%表格

\usepackage{threeparttable}  %表格注释
\usepackage[pdfstartview=FitH,
CJKbookmarks=true,
bookmarksnumbered=true,
bookmarksopen=true,
colorlinks, 
pdfborder=001,
linkcolor=blue,
anchorcolor=blue,
citecolor=blue
]{hyperref}

\newtheorem{theorem}{Theorem}

\theoremstyle{definition}

\begin{document}
	
%\linenumbers

\title{The Shape of Information: Global Information Geometric Limits in Multi-task Quantum Systems}%

\author{Zishuo Ren}%
\affiliation{Beijing Key Laboratory of Quantum Sensing and Precision Measurement, and Center for Quantum Information Technology, and Institute of Quantum Electronics, Peking University, Beijing 100871, China}
%\author{Yiting Lin}%
%\affiliation{Beijing Key Laboratory of Quantum Sensing and Precision Measurement, and Center for Quantum Information Technology, and Institute of Quantum Electronics, Peking University, Beijing 100871, China}
%\author{Zimeng Li}%
%\affiliation{Beijing Key Laboratory of Quantum Sensing and Precision Measurement, and Center for Quantum Information Technology, and Institute of Quantum Electronics, Peking University, Beijing 100871, China}
%\author{Zhuochen Li}%
%\affiliation{Beijing Key Laboratory of Quantum Sensing and Precision Measurement, and Center for Quantum Information Technology, and Institute of Quantum Electronics, Peking University, Beijing 100871, China}

\author{Ziyang Chen}%
\email[E-mail: ]{chenziyang@pku.edu.cn}
\affiliation{Beijing Key Laboratory of Quantum Sensing and Precision Measurement, and Center for Quantum Information Technology, and Institute of Quantum Electronics, Peking University, Beijing 100871, China}
\affiliation{State Key Laboratory of Environment Characteristics and Effects for Near-space, Center for Photonic Quantum Precision Measurement, Advanced Research Institute of Multidisciplinary Science, Beijing Institute of Technology, Beijing 100081, China}

\author{Hong Guo}%
\email[E-mail: ]{hongguo@pku.edu.cn}
\affiliation{Beijing Key Laboratory of Quantum Sensing and Precision Measurement, and Center for Quantum Information Technology, and Institute of Quantum Electronics, Peking University, Beijing 100871, China}

%\date{September 2022}%
%\date{\today}%

\begin{abstract} 
Future quantum networks are expected to perform multiple tasks simultaneously within a single system, such as integrated sensing and communication (ISAC) architectures. %multiple tasks including secure communication, precise sensing and distributed computing under limited available quantum resources. Despite the ongoing emergence toward experimental architectures like integrated sensing and communication (ISAC), it is unclear how much capacity can a physical carrier provide and what tradeoff constrains the simultaneous support of multiple tasks. 
%By identifying the information geometry of shared quantum carriers, we have found the underlying object governing the capacity and allocability of multi-task systems: the global quantum fisher information matrix (g-QFIM), and yielded a Shannon capacity like bound. 
Despite various metrics, we find that the evaluation of multiple tasks can be unified by their information capacity, and the total task capacity is not determined simply by addition, but is fundamentally constrained by an information geometry which we call the global quantum Fisher information matrix (g-QFIM). With this insight, we derive a non-asymptotic, measurement-independent upper bound on the Holevo information for multi-task systems, which takes a Shannon-capacity-like form. %太长
It not only quantifies the capacity limit, but also the allocability. Our results reveal a structural phase transition in multi-task performance under resource variation, where additional physical resources no longer increase independent task capacity but instead concentrate more information into a new mono-task mode. Numerical simulations based on photonic phase encoding and realistic noise channels confirm these predictions.
This work establishes a unified information-geometric principle for quantum multi-task systems, with implications for the design of future quantum networks and ISAC architectures.
%The result shows that this bound clearly reveals the fundamental tradeoff and structure transition in multi-task systems, providing the principle for multi-task system design and suggesting a profound geometry underlying the limit of information acquisition.

%By combining Holevo information with directional quantum Fisher information, we derive two upper bounds: a task-allocation bound that quantifies the tradeoff among different task directions, and a total information bound governed by the global quantum Fisher information matrix. These results connect global information acquisition with local distinguishability geometry, yielding a computable, measurement-independent ceiling that remains applicable beyond asymptotic regimes. Numerical simulations show how physical resources and noise jointly shape both the total information capacity and its allocability. The resulting structure parallels the geometric form of Shannon-type capacity bounds, suggesting a broader geometric principle underlying the limits of quantum information acquisition.

\end{abstract}

\maketitle
%\tableofcontents  %目录

%\section{INTRODUCTION}\label{sec1}

{\textit{Introduction}}---
Future quantum networks~\cite{azuma2023quantum,simon2017towards,kimble2008quantum} and integrated sensing-communication (ISAC) systems~\cite{liu2024quantum,liu2024integrated,lu2024integrated,liu2022integrated,chen2022quantum,marra2018ultrastable} are inherently multi-task platforms~\cite{wei2022towards,awschalom2021development,wehner2018quantum}, where a single physical carrier can be simultaneously used for communication~\cite{gisin2007quantum}, sensing~\cite{degen2017quantum}, computing~\cite{ladd2010quantum} and other tasks that aim to acquire and process information. This raises a basic question: What is the fundamental limit on the total information that a single physical carrier can support when multiple tasks compete for the same quantum resources?

%由量子理论催生的三大量子技术（量子计算、量子通信、量子传感）正在不断扩展我们在理解、模拟、甚至改造自然方面的边界。这些丰富的应用场景中的每一类都需要特定的量子资源以及相应的评价指标。然而，具有高相干性乃至复杂纠缠结构的量子资源的大规模制备和长时间存储极大地制约着量子技术的使用场景。因此，未来量子信息系统的目标并不只是增大单个量子节点所能容纳的量子资源，更需要发展有限的量子资源下能够同时支撑安全通信、分布式计算、时间同步以及高精度传感的多功能量子信息理论与技术，而这也正是量子网络/量子互联网的初衷所在。因此，一个本质的问题自然出现：当受限量子物理载体被多个任务共同复用时，如何衡量系统的性能？
%The thrive of quantum information theory has given birth to many revolutionary technologies such as quantum sensing~\cite{degen2017quantum}, quantum computing~\cite{ladd2010quantum} and quantum communication~\cite{gisin2007quantum}, each requiring the corresponding quantum resources.

%However, high-quality quantum resources remain difficult to generate, distribute, and preserve on a scale, making isolated task-specific implementations increasingly inefficient. 
%Future quantum networks thus require not only more resources to manipulate and distribute at scale~\cite{azuma2023quantum,simon2017towards,kimble2008quantum}, but also principled ways to reuse physical resources across communication, computing, timing, and sensing tasks~\cite{wei2022towards,awschalom2021development,wehner2018quantum}.
%Therefore, a fundamental question naturally arises: How do we measure the performance of a system shared by multiple tasks?

%以两种典型的任务：通信和传感为例，近年来关于多任务复用性能的工程优化已在经典电磁波技术中形成了ISAC的概念[引综述]，并逐渐向量子领域扩展[引清华和山西大学那几个]。然而，为通信和传感分别选取工程指标后画出经验的tradeoff曲线这种ad hoc的处理方式只是治标不治本，解决这个问题真正需要的是一种根本、普适的约束。对多任务之间共同点的认识不足导致这一问题仍是亟待解决的开放问题[引问题综述]。
%Take two typical tasks, commmunication and sensing, for example. 

Existing works typically evaluate multi-task performance using task-specific metrics and ad hoc optimization~\cite{lu2024integrated}. However, these methods either rely on additional assumptions such as asymptotic assumption, or follow single-task formulations that do not capture the global structure of information sharing across tasks. As a result, a unified and universal principle governing multi-task systems remains missing, and has become a pressing open challenge~\cite{zhou2025towards,lu2024integrated,giani2024quantum}.

%In recent years, the study of multi-task systems has formed the concept of ISAC in classical electromagnetic wave technology~\cite{lu2024integrated,liu2022integrated}, and similar ideas now begin to emerge in quantum settings~\cite{liu2024quantum,liu2024integrated,chen2022quantum,marra2018ultrastable}. Despite engineering progress, performance limits and tradeoffs are still studied through ad hoc optimization, while the common information-theoretic structure underlying the tasks remains insufficiently understood. Establishing such a structure has therefore become a pressing open challenge in both fields~\cite{lu2024integrated,zhou2025towards}.

\begin{figure*}%[t]
	\centering
	\includegraphics[width= 0.9\linewidth]{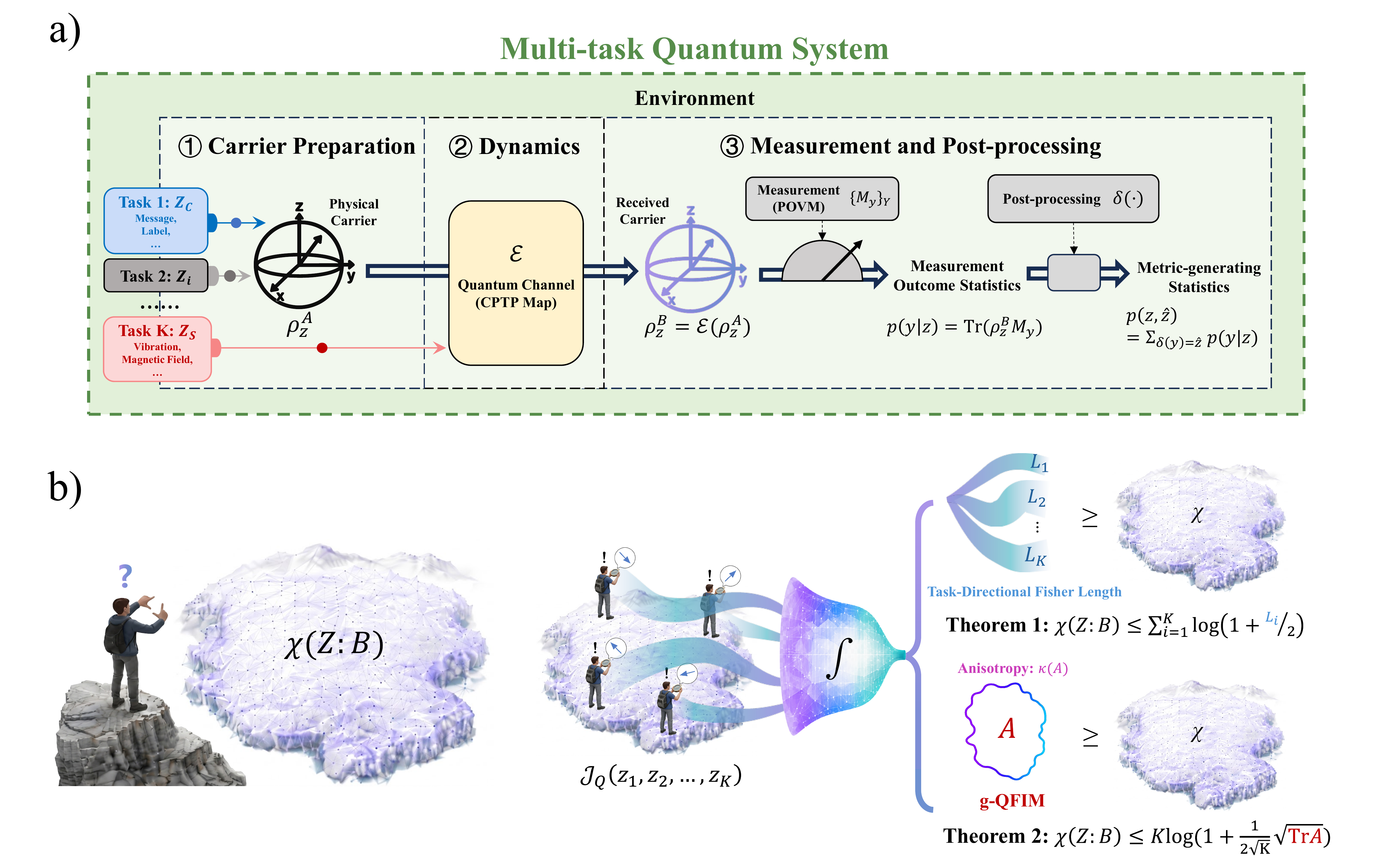}
	\caption{Panel a): The 3-step common information-acquisition structure shared by many tasks; Panel b): The visualization of our two theorems which first bound the global information capacity with local fisher information and then discover the vital role the shape of information plays in multi-task systems.}\label{unified scheme}
    \end{figure*}

In this work, we find that although the metrics of different tasks vary, they share a similar information structure~\cite{nielsen2010quantum} and the same physical carrier in multi-task systems (Fig~\ref{unified scheme}). Under this structure, the performance of multiple tasks is unified by the Holevo information capacity, which is not determined simply by adding up single-task capacities, but constrained by an underlying information geometry of the physical carrier. This geometry is captured by a global quantum Fisher information matrix (g-QFIM), which represents the average sensitivity of the quantum state to certain parameters within the given prior region. With this insight, we derive two Shannon-capacity-like~\cite{shannon1948mathematical} upper bounds on the Holevo information for multi-task systems as two main theorems: the first is a capacity upper bound on each task, which visualizes and indicates the fundamental capacity tradeoff between tasks; the second is a further upper bound, which not only gives a total capacity bound, but also quantifies the allocability of systems. 

As a consequence, we find that the ``shape" of information plays a significant role in multi-task systems. Specifically, our results reveal a previously unrecognized structural phase transition in multi-task systems: increasing physical resources initially enhances independent task capacities, but beyond a critical point, information becomes increasingly concentrated into collective modes for less tasks, reducing effective task separability. Numerical simulations based on photonic phase encoding and several typical noises confirm and further demonstrate our prediction. Therefore, this work establishes a unified and universal framework for multi-task systems and indicates the profound role of information geometry in the design of future quantum networks and ISAC systems.

\textit{Problem setting}---As shown in Fig~\ref{unified scheme}, we consider a general quantum multi-task system characterized by task variables \(Z\), which the receiver is required to infer, estimate, discriminate, classify, or decode. Their subscripts indicate the corresponding tasks.

The central observation behind our formulation is that a broad class of physical tasks share the same 3-step information-acquisition structure. First, the physical carrier $\rho^A_z$, into which $Z$ are encoded, is prepared. Second, the carrier goes through a quantum channel $\mathcal{E}$, representing evolution, process or propagation of the carrier. For communication tasks, $Z_C$ are encoded earlier in the first step; for sensing tasks, $Z_S$ are encoded later in the evolution step; while for computing tasks $Z$ can be encoded in both steps. Third, the receiver will receive $\rho_z^B$ and perform a POVM \(\mathcal M=\{M_y\}_Y\) with $M_y\ge0$ and $\Sigma_y M_y=I$, producing a classical observation $y$ which satisfies $p(y|z)=\operatorname{Tr}\!\left(\rho_z^B M_y\right)$, and the final output of the task, $\widehat Z$, is obtained through a post-processing rule $\delta(\cdot)$: $\widehat Z=\delta(Y)$, where \(\hat Z\) may be an estimate, a hypothesis decision, a class label,
or a decoded message, depending on the task.

Before measurement $\{M_y\}_Y$ and processing rule $\delta$ are specified, the only task-independent quantity at the receiver is the classical information encoded in the quantum ensemble $\{p(z),\rho_z^B\} $ itself, quantified by the Holevo information $\chi(Z:B)$~\cite{holevo1973bounds}. Once $M$ and $\delta$ are fixed, the task output $\widehat Z$ gives only a downstream, measurement-dependent information $I(Z;\widehat Z)$ that summarizes the same distribution $p(z,\widehat z)=p(z)\sum_{y:\delta(y)=\widehat z}\operatorname{Tr}(\rho_z^B M_y)$ from which specific metrics such as estimation error, detection probability or decoding error are evaluated. According to the data processing inequality and the Holevo bound:
\begin{equation}
    I(Z;\widehat Z)
    \le
    I(Z;Y)
    \le
    \chi(Z:B).
\end{equation}
Therefore, $\chi(Z:B)$ is the common and fundamental quantity for characterizing information-theoretic limits.

As the first and the most visible step towards our main theorems, we consider two tasks, each encoded in $\rho^B_z$ by a scalar continuous task variable $Z_i$ with a commonly used finite prior support $\{[a_i,b_i]\}$: 
\begin{equation}
    Z=(Z_1,Z_2),
    \qquad
    Z_i\in[a_i,b_i],
    \qquad
    i=1,2 ,
\end{equation}
For clarity, we first consider tasks whose variables are independent of each other due to different utility, and the correlated prior cases are further given in Appendix D. The prior distribution $p_i$ are written as:
\begin{equation}
    p(z_1,z_2)
    =
    p_1(z_1)p_2(z_2),
    \qquad
    \operatorname{supp}p_i\subset[a_i,b_i].
\end{equation}
 Next we will see how $Z$ and $\rho_z^B$ shapes $\chi(Z:B)$, the capacity of multi-task systems.

%Before any task-specific measurement and processing rule are fixed, the received quantum ensemble is characterized by a more fundamental information budget: the classical information about the task variable encoded in the quantum states. The Holevo information quantifies this measurement-independent budget, while concrete task metrics arise only as downstream operational manifestations.

%At the level of the received quantum ensemble, the common object is the amount of classical information about the task variable that can be extracted by any downstream procedure. For any measurement and any subsequent decision rule, the data-processing principle and the Holevo information give:
%\begin{equation}
%  I(Z;\widehat Z)
%   \le
%   I(Z;Y)
%   \le
%    \chi(Z:B).
%\end{equation}
%Thus, the Holevo information is used as a measurement-independent global information budget constraining the performance of all information-acquisition tasks.

\textit{Main theorem}---Based on the above problem setting, we now present our two main theorems. The detailed proof can be found in the Appendices.

\begin{theorem}[Separable capacity bound]
\label{thm:two-task-bound}
This theorem shows how each task contributes to the total capacity. The total capacity is constrained by the average sensitivity of $\rho_z^B$ to two task variables:
\begin{equation}
    \chi(Z_1,Z_2,...,Z_K:B)
    \le
    \sum_{i=1}^{K}
    \log
    \left(
    1+\frac{L_i}{2}
    \right),~K=2.
    \label{eq:two-task-holevo-bound}
\end{equation}
\end{theorem}
When $K=2$, the quantities \(L_1\) and \(L_2\) are task-directional Fisher lengths and they are defined as:
\begin{equation}
    L_i
    :=
    \int_0^1\sqrt{F_i(\phi_i)}\,d\phi_i,~i=1,2
    \label{eq:two-task-lengths}
\end{equation}
where
\begin{equation}
\begin{aligned}
    &F_1(\phi_1)
    :=
    \int_0^1
    \widetilde p_2(\phi_2)
    \widetilde{\mathcal J}_{Q,11}(\phi_1,\phi_2)
    \,d\phi_2 .\\
    &F_2(\phi_2)
    :=
    \int_0^1
    \widetilde p_1(\phi_1)
    \widetilde{\mathcal J}_{Q,22}(\phi_1,\phi_2)
    \,d\phi_1 .
\end{aligned}
    \label{eq:F-two-task}
\end{equation}
$\phi_i$, $\widetilde p_i$ and $\widetilde{\mathcal{J}}_Q$ correspond to the normalized $Z_i$, $p_i$ and QFIM ${\mathcal{J}}_Q$.

\begin{theorem}[Information-geometric capacity bound]
\label{thm:two-task-trace-bound}
This theorem shows what underlying object governs the capacity of multi-task systems. The total capacity is fundamentally constrained by g-QFIM $A$:
\begin{equation}
    \chi(Z_1,Z_2,...,Z_K:B)
    \le
    K\log\!\left(
    1+\frac{1}{2\sqrt K}
    \sqrt{\operatorname{Tr}A}
    \right).
    \label{eq:two-task-trace-bound}
\end{equation}
\end{theorem}
When $K=2$, the g-QFIM $A$ is defined as:
\begin{equation}
    A
    :=
    \int_{[0,1]^2}
    R(\boldsymbol\phi)
    \widetilde{\mathcal J}_Q(\boldsymbol\phi)
    R(\boldsymbol\phi)
    \,d\boldsymbol\phi ,
    \label{eq:two-task-effective-A}
\end{equation}
where 
\begin{equation}
\begin{aligned}
    R(\boldsymbol\phi)
    &:=
    \operatorname{diag}
    \left(
    \sqrt{w_1(\boldsymbol\phi)},
    \sqrt{w_2(\boldsymbol\phi)}
    \right)\\
    &=
    \operatorname{diag}
    \left(
    \sqrt{\widetilde p_2(\boldsymbol\phi)},
    \sqrt{\widetilde p_1(\boldsymbol\phi)}
    \right).
\end{aligned}
\end{equation}

Thm.~\ref{thm:two-task-bound} and Thm.~\ref{thm:two-task-trace-bound} follow a natural way to characterize $\chi(Z:B)$, which lacks a tractable expression, and is generally difficult to compute and optimize analytically in experiments~\cite{gyongyosi2018survey,holevo2012quantum,roga2010universal,beigi2007complexity}: to bound this global quantity by calculable local quantities. Several fundamental results follow this local-to-global route as well: the I-MMSE relation~\cite{guo2005mutual} directly connects information gain with experimentally meaningful quantities, but it describes a specific task under a given operating condition; Efroimovich~\cite{efroimovich1980information}- and van Trees~\cite{van2004detection,bj/1186078362}-type inequalities reveal the Fisher-information structure behind global performance limits and admit multi-parameter formulations, but they usually rely on regular priors or asymptotic conditions; More recently, Górecki inequalities~\cite{gorecki2025mutual} take a great leap by providing non-asymptotic Fisher-information bounds that are robust to finite supports and non-smooth priors, but their proof essentially relies on a one-dimensional Sobolev embedding which fails in higher dimensions~\cite{brezis2011functional}, and thus the result cannot be directly applied to multi-task systems.

To derive a bound that is simultaneously non-asymptotic, robust to finite priors, and compatible with high-dimensional structure to meet the need in quantum multi-task systems, we inherit the advantages of the above results to construct our two main theorems. First, we borrow the idea of directly combining physical resources with information-theoretic limits; Second, we retain the high-dimensional generalization ability of van Trees-type inequalities~\cite{bj/1186078362}; and third, we absorb the robustness of Górecki inequalities to finite support, non-smooth priors, and non-asymptotic conditions. 

%thm1的解释

%The Fisher length $L_1$ and $L_2$ describe how local information on the task direction accumulates to bound the global information $\chi(Z_1,Z_2:B)$. They quantify how much local QFIM strength can be accumulated into a global information budget along each task direction. 

Thm.~\ref{thm:two-task-bound} gives a separable interpretation of the multi-task information carried by a quantum ensemble. The quantity $\chi(Z_1,Z_2:B)$ represents the total classical information about the two task variables available at the receiver before any measurement is chosen. And $L_1$, $L_2$ describe how much information is accumulated through local distinguishability on each task variable. %Through Thm.~\ref{eq:F-two-task}, we are able to see how the local property of the quantum ensemble gives the information-theoretic bound on the performance of each task.
Specifically, for a fixed point $\phi=(\phi_1,\phi_2)$ in the task variable space, the QFIM element $\widetilde J_{Q,ii}(\phi)$ measures the infinitesimal distinguishability of the received state $\rho_z^B$ under a change of the $i$-th task coordinate. Thus, it quantifies how strongly the physical carrier responds locally to that task variable. Since the other task variable is not fixed, its contribution is averaged over its prior distribution. The remaining integral over $\phi_i$ then accumulates this local response over the finite support of the $i$-th task variable. In this sense, $L_i$ measures how many locally resolvable parameter intervals fit into the prior support, or in other words, represents the best ``signal-to-noise ratio (SNR)"~\cite{shannon1998communication} when using $\rho^B_{\phi}$ to distinguish $\phi_i$ in $[0,1]$.

According to Thm.~\ref{thm:two-task-bound}, the joint Holevo information cannot exceed the sum of the two directional bounds $\log(1+L_i/2)$. A large $L_i$ means that the received states trace a long distinguishable trajectory when the $i$-th task variable changes, so the carrier can in principle support more information about that task. Conversely, if $L_i$ is small, the states remain nearly indistinguishable along that direction, and no measurement can extract much task information from it. Therefore, a carrier with a large $L_1$ but a small $L_2$ is not genuinely balanced for two-task reuse, even if its total local distinguishability is large. Balanced multi-task support requires the information geometry of the received ensemble to allocate comparable information along both task directions.

Thm.~\ref{thm:two-task-bound} already indicates that information in the multi-task systems is not merely a scalar quantity, but a structural property of the received ensemble instead: it depends not only on the amount of information, but also on the ``shape" of information. However, Thm.~\ref{thm:two-task-bound} exposes this structure only through directional lengths $L_i$. Although $L_i$ is useful to show the intrinsic tradeoff between prescribed task variables, they do not retain the full information structure of the quantum ensemble, such as correlations between task variables or the possible concentration of information into lower-dimensional subspaces.%, or the invariance under reparameterization.

To uncover the complete structure, we further derive the Thm.~\ref{thm:two-task-trace-bound}. This leads to the g-QFIM $A$, which captures the whole shape of information and allows us to discuss the capacity and allocability respectively.

%thm2的解释

Thm.~\ref{thm:two-task-trace-bound} indicates that the capacity is completely constrained by the trace of $A$, which is exactly the underlying structure we are looking for. %The trace gives the coarse total ceiling, whereas the off-diagonal entries and the spectrum of \(A\) retain how the two task directions are coupled by the same local QFIM geometry.
%The total theorem shows that the joint Holevo information of the two scalar tasks is controlled by the coordinate-free scalar
%\begin{equation}
%    \mathcal T
%    =
%    \sum_{i=1}^{2}
%    \int
%    \widetilde{\mathcal J}_Q(\mathsf D_i,\mathsf D_i)
%    \,d\mu_i.
%\end{equation}
%This quantity measures the total amount of task-weighted local distinguishability accumulated along the two task directions. In the adapted normalized coordinates, the same scalar becomes
%\begin{equation}
%    \mathcal T=\operatorname{Tr}A,
%\end{equation}
%with
%\begin{equation}
%    A=
%    \int
%    R\widetilde{\mathcal J}_Q R\,d\boldsymbol\phi.
%\end{equation}
%In other words, the trace contains more than a upper bound for $\chi(Z_1,Z_2:B)$. It also demonstrates the allocability and the capacity, respectively, by showing the information structure of $A$. 
Since the trace can be written as:
\begin{equation}
    \operatorname{Tr}A=\sqrt{\operatorname{det}(A)}\left(\sqrt{\kappa(A)}+\dfrac{1}{\sqrt{\kappa(A)}}\right),
\end{equation}
where $\operatorname{det}(A)$ stands for the capability of the multi-task system and $\kappa(A)=\lambda_{\operatorname{max}}(A)/\lambda_{\operatorname{min}}(A)$ describes the imbalance between two information axes, thus shaping the information allocation between tasks.
%Its diagonal entries are the accumulated effective QFIM strengths along the two task directions. Its off-diagonal entries quantify whether the two task directions are locally coupled through the same information geometry. Thus, two systems may have the same total ceiling \(\mathcal T\) but very different task structures.
For example, a nearly isotropic spectrum of \(A\) indicates balanced support across the task directions. A strongly anisotropic or nearly low-rank spectrum indicates that the available information is concentrated along a few coupled task combinations. The tradeoff between tasks originates from the fact that: a single physical carrier defines an information manifold, and different task variables correspond to different directions on this manifold. Therefore, resources do not increase all directions equally. Also,
information is redistributed rather than simply increased.%Therefore, \(\mathcal T\) gives the coarse total joint ceiling, while the full matrix \(A\) diagnoses how that total information is organized across task directions.

%The theorem remains a joint-information upper bound. It does not by itself quantify separated-task nuisance penalties. Such penalties depend on finer task-specific structures, such as conditional Holevo information, effective Fisher information, or Schur-complement-type quantities. The off-diagonal structure and spectrum of \(A\) provide diagnostics for when such penalties may become significant, but the total theorem itself only bounds the joint Holevo information.

%The preceding results establish a non-asymptotic bridge from local QFIM geometry to global Holevo information.  We now clarify how this theorem should be read before turning to numerical examples.  The purpose of the simulations below is not to prove the bound again, but to illustrate what kind of physical structure is revealed once the local matrix $\mathcal J_Q$ is lifted to task-level global information budgets.

%For a rectangular task prior
%\[
%\Theta_k\in[a_k,b_k],\qquad W_k=b_k-a_k,
%\]
%we use the dimensionless effective information matrix
%\[
%A := D\,\mathcal J_Q\,D,
%\qquad
%D=\operatorname{diag}(W_1,W_2).
%\]
%For a constant QFIM over the prior region, the task-directional
%Fisher lengths reduce to
%\[
%L_k=\sqrt{A_{kk}},
%\qquad
%B_k:=\log(1+L_k/2),
%\]
%where $B_k$ denotes the one-dimensional non-asymptotic ceiling
%associated with the $k$th task direction.  These quantities should not
%be interpreted as the separately achievable performance of two
%independently optimized tasks.  Rather, %they are directional global
%ceilings induced by the same joint quantum statistical model.  In this
%sense, the curve

Therefore, we are able to reveal three levels of the theoretical problems in multi-task system. First, the total
information capacity layer controlled by $\operatorname{Tr} A$,
%and
%\[
%B_{\rm tot}(T)
%=
%2\log\!\left(1+\frac{\sqrt{T}}{2\sqrt{2}}\right),
%\]
which does not care how information is distributed and only gives an upper bound on how much classical information can be obtained through multiple tasks when given the physical carrier and prior distribution.
Second, the structural layer controlled by allocability $\kappa(A)$ which
diagnoses whether the geometry of the information is isotropic or
anisotropic, and capability $\operatorname{det}(A)$ that determines whether the available information is distributed across independent directions or concentrated into a lower-dimensional structure.% A nearly balanced spectrum indicates that the carrier supports two comparable task directions, whereas a highly anisotropic or nearly rank-deficient spectrum suggests that the system effectively favours a dominant combination of task parameters. 
Third, the task layer controlled by the directional information budgets $(L_1,L_2)$, which describe how the information is actually exposed and distributed along the task axes. Thm.~\ref{thm:two-task-trace-bound} solves the first and second layer, while Thm.~\ref{thm:two-task-bound} solves the third layer.

\textit{Simulation setup}---To make the implications of Thm.~\ref{thm:two-task-bound} and Thm.~\ref{thm:two-task-trace-bound} more explicit, %First, how does the total QFIM-induced amount $T$ track the true Holevo information in the weak-light regime?  Second, how does ideal resource allocation reshape the two task-directional ceilings?  Third, how do physical noise channels deform not only the magnitude, but also the spectrum and local exchange structure of the task-budget curve? 
we consider a minimal model: two-mode quantum carrier with vacuum and two
single-photon polarization modes, $\{|0\rangle,|H\rangle,|V\rangle\}$.
The task variables, either a field strength to be sensed or a message to be encoded, are encoded in two phases $\boldsymbol\theta=(\theta_H,\theta_V)$ of the carrier
\begin{equation}
    |\psi_{\boldsymbol\theta}\rangle
=
\sqrt{1-N}\,|0\rangle
+
\sqrt{n_H}\,e^{i\theta_H}|H\rangle
+
\sqrt{n_V}\,e^{i\theta_V}|V\rangle,
\end{equation}
where $n_H+n_V=N,~ n_H=\eta N,~n_V=(1-\eta)N$.

Here $N$ is the mean excitation in the vacuum--single-photon model, representing the physical resource used in multi-task systems, and $\eta\in[0,1]$ represents the physical resource-allocation parameter
between the two tasks.

We will visualize the previous implications by changing $N$ and use
\[
\eta \mapsto (L_1(\eta),L_2(\eta))
\]
to give task tradeoff curves. Under different non-ideal effects, the curve family will demonstrate the overall evaluation of multi-task performance.

%This separation is important for interpreting ``allocability''.  In what follows, we use allocability in a local operational sense: it refers to how effectively a physical resource-allocation parameter changes the relative task budgets near a given operating point.  A natural local descriptor is the marginal exchange rate
%\[
%\Gamma(\eta)
%=
%-\frac{dB_2}{dB_1}
%=
%-\frac{dB_2/d\eta}{dB_1/d\eta},
%\]
%which measures the local budget cost of transferring information from one task direction to the other.  This may be complemented by a task-share response
%\[
%s(\eta)=\frac{B_1(\eta)}{B_1(\eta)+B_2(\eta)},
%\qquad
%\mathcal A_{\rm resp}(\eta)=\left|\frac{ds}{d\eta}\right|.
%\]
%Thus the global span of a plotted curve is only a secondary descriptor: the more intrinsic notion is how sensitively and at what marginal cost the task-budget ratio changes under a physically meaningful resource control.

For the pure state defined above, the QFIM with respect to
$(\theta_H,\theta_V)$ is
\begin{equation}
    \mathcal J_Q
=
4
\begin{pmatrix}
n_H(1-n_H) & -n_H n_V\\
-n_H n_V & n_V(1-n_V)
\end{pmatrix}.
\end{equation}

We take a rectangular phase prior with $
W_H=W_V=2\pi,~D=2\pi I_2$, so that $A=D\mathcal J_QD$.

%All numerical illustrations are restricted to the weak-excitation region $N<1/2$.  This is not a mathematical requirement of the theorem, but a physical validity condition of the vacuum--single-photon truncated model.  In this regime, the vacuum component remains a non-negligible phase reference, and for every allocation $\eta$ both mode occupations satisfy $n_H,n_V<1/2$, so the directional QFIs $4n_i(1-n_i)$ remain in their monotonic small-signal region.  Beyond this regime, the truncated model approaches a different high-excitation limit in which the vacuum reference is depleted and the two absolute phase tasks gradually collapse toward a relative-phase problem.

For the full $2\pi$ phase prior, the ensemble average is diagonal,
\begin{equation}
\bar\rho
=
(1-N)|0\rangle\langle0|
+
n_H|H\rangle\langle H|
+
n_V|V\rangle\langle V|,    
\end{equation}
and we use the noiseless Holevo information $\chi_{\rm true}=S(\bar\rho)=H(1-N,n_H,n_V)$, as a benchmark for comparison.

\textit{Non-ideal effect models}---To show how physical non-ideal effects reshape the information geometry and task tradeoff curves, we consider three noise mechanisms. All are applied at the quantum state level, and the resulting mixed-state QFIM is computed by the symmetric logarithmic derivative (SLD) formula
\begin{equation}
[\mathcal J_Q]_{ij}
=
2\mathrm{Re}\sum_{\lambda_\mu+\lambda_\nu>0}
\frac{
\langle \mu|\partial_i\rho|\nu\rangle
\langle \nu|\partial_j\rho|\mu\rangle
}{
\lambda_\mu+\lambda_\nu
},    
\end{equation}

where
$
\rho
=
\sum_\mu \lambda_\mu |\mu\rangle\langle \mu|.
$
\paragraph{Attenuation.}
The attenuation effect can be modeled by gradually changing $N$ from $1$ to $0$.

\paragraph{Phase diffusion.}
The random phase jitter can be modeled as
$
\theta_H\mapsto\theta_H+\delta_H,
~
\theta_V\mapsto\theta_V+\delta_V,
$
where $(\delta_H,\delta_V)$ are zero-mean Gaussian random variables.
For the symmetric case used in the main figure,
$
\sigma_H=\sigma_V=\sigma_\phi,
~
\operatorname{corr}(\delta_H,\delta_V)=0.
$
At $\theta_H=\theta_V=0$, the phase-diffused density matrix has the same
populations as the pure state, while the coherences are damped as
$
\rho_{0H}\mapsto e^{-\sigma_H^2/2}\rho_{0H},
\qquad
\rho_{0V}\mapsto e^{-\sigma_V^2/2}\rho_{0V},
$
and
\begin{equation}
    \rho_{HV}
\mapsto
\exp\!\left[
-\frac12
\left(
\sigma_H^2+\sigma_V^2
-2c\,\sigma_H\sigma_V
\right)
\right]\rho_{HV},
\end{equation}
where $c=\operatorname{corr}(\delta_H,\delta_V)$.  This noise primarily
degrades phase coherence and therefore directly attacks the  tasks.

\paragraph{Unresolved mode crosstalk.}
The mode-mixing noise channel acting on the
$\{|H\rangle,|V\rangle\}$ subspace can be modeled through a CPTP channel $\mathcal{E}$~\cite{nielsen2010quantum}. A known deterministic mode rotation would be a parameter-independent unitary and would not reduce the QFIM.
To model crosstalk, we therefore use an unresolved random-unitary channel
\begin{equation}
    \mathcal E_\alpha(\rho)
=
\frac12 U(+\alpha)\rho U^\dagger(+\alpha)
+
\frac12 U(-\alpha)\rho U^\dagger(-\alpha),
\end{equation}
where
\begin{equation}
\begin{aligned}
    U(\alpha)
=
|0\rangle\langle0|
&+
\cos\alpha
\left(
|H\rangle\langle H|+|V\rangle\langle V|
\right)\\
&+
\sin\alpha
\left(
|V\rangle\langle H|-|H\rangle\langle V|
\right).
\end{aligned}
\end{equation}

This channel represents unresolved polarization or mode scrambling. It does not merely attenuate the total QFIM amount; it can also strongly reshape $A$ by mixing the physical channels through which task information is carried.

\textit{Results}---
First, we show how our total upper bound tracks the true Holevo information.
\begin{figure}[h]
    \centering
    \includegraphics[width=0.72\linewidth]{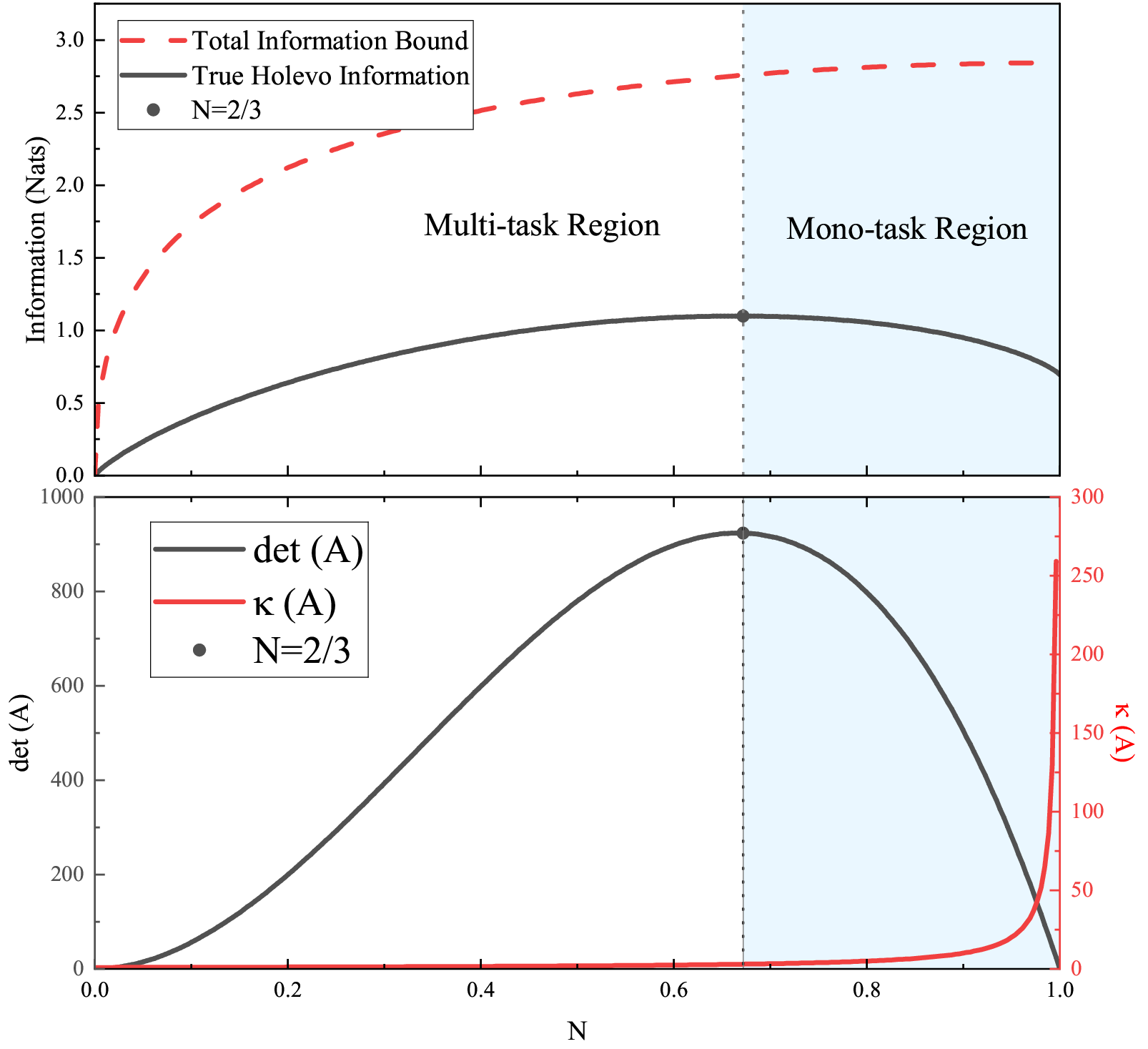}
    \caption{
    Total Holevo information and QFIM-induced ceilings. In the upper panel, the solid curve shows the exact Holevo information of the noiseless phase ensemble and the dashed curve shows
    the trace-based total information bound; In the lower panel, the transition of information structure is shown through $\det(A)$ and $\kappa(A)$.% The comparison is intended as a global-local consistency check, not as an achievability claim.
    }
    \label{fig:total_ceiling}
\end{figure}

Figure~\ref{fig:total_ceiling} compares the true Holevo information
$\chi_{\rm true}$ of the noiseless phase ensemble with the upper bound~\eqref{thm:two-task-trace-bound} when the physical resource $N$ changes. 
%and the second is the task-resolved sum
%\[
%B_{\rm task}
%=
%\log(1+L_1/2)+\log(1+L_2/2).
%\]
Within the weak-excitation
regime ($N\le1/2$), the local QFIM geometry gives a sensible non-asymptotic ceiling
for the global information carried by the physical carrier. %The figure is not meant to claim achievability or tightness of the bound.  Rather, it serves as a sanity check for the global-local information link.
When $N\ge2/3$, we can clearly see that $\operatorname{det}(A)$ starts to decrease, showing that additional physical resource no longer increases the two-dimensional independent information volume. Instead, the extra local distinguishability is increasingly concentrated in the relative-phase mode $\theta_H-\theta_V$, as also reflected by the growing spectral anisotropy $\kappa(A)$. 

Thus, $N=2/3$ is identified by our theory as the optimal resource point for multi-task: beyond it, the carrier still accumulates more useful information budget, but this information budget is organized in a structure better matched to a collective parameter than to two independent tasks. Correspondingly, $\chi_{\rm true}$ starts to decrease when $N\ge2/3$, indicating a transition of the information structure. For general systems, we provide a criterion that this optimal resource point may be found when $\left.
\frac{\partial \det A}{\partial N}
\right|_{N=N_*}
=0,
~
\left.
\frac{\partial^2 \det A}{\partial N^2}
\right|_{N=N_*}
<0,
~
\left.
\frac{\partial \kappa(A)}{\partial N}
\right|_{N=N_*}
>0$.

To further illustrate how different physical imperfections affect
multi-task performance, we track the task tradeoff under three typical noises, and the three mechanisms produce qualitatively distinct geometric responses.

\begin{figure}[h]
    \centering
    \includegraphics[width=\linewidth]{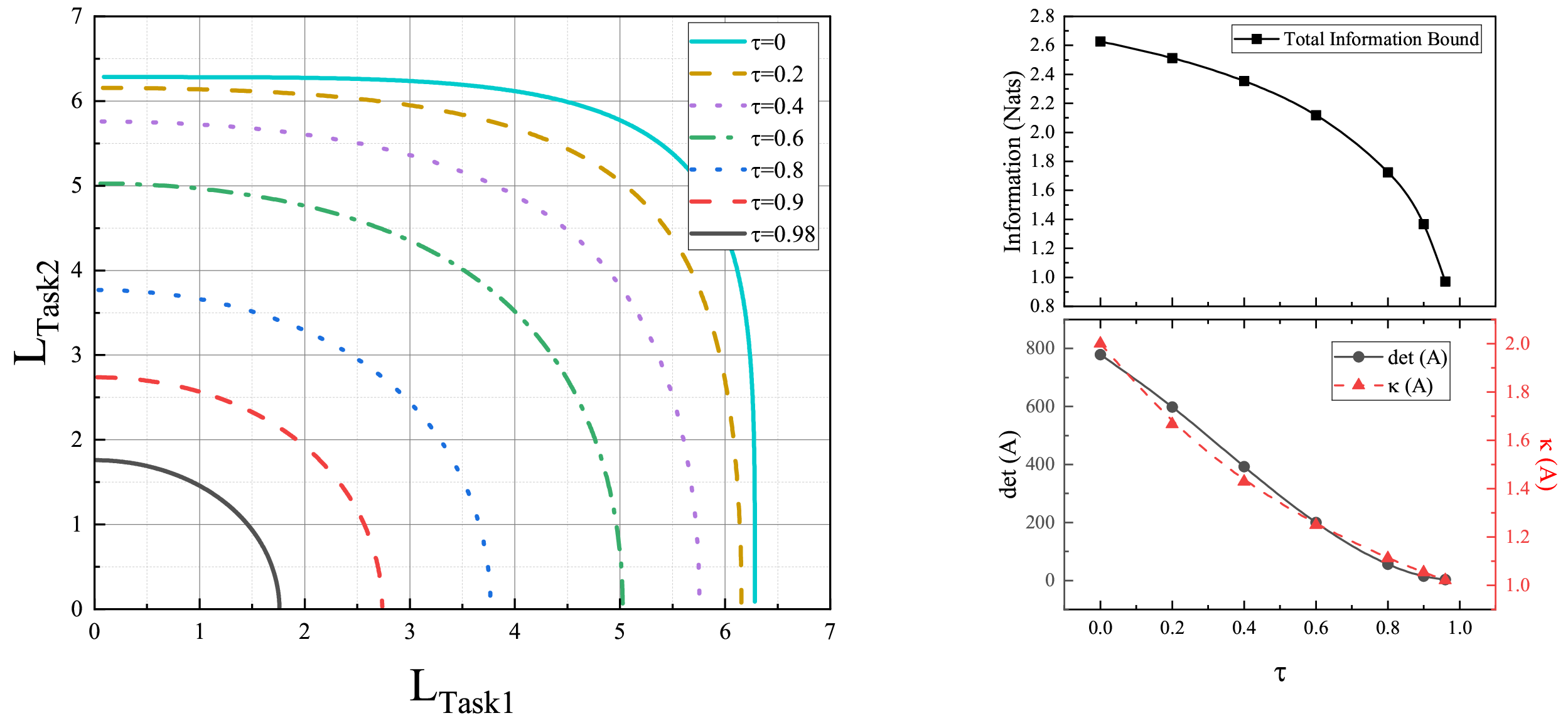}
    \caption{
    Tradeoff curve family under attenuation. The curve corresponds to the attenuation factor $\tau=1-N/N_0$ gradually increasing from $0$ to $0.98$ at $N_0=1/2$. Left: the tradeoff curve family; Right top: the total bound derived by $\mathrm{Tr}A$ at $\eta=1/2$; Right bottom: $\det(A)$ and $\kappa(A)$ at $\eta=1/2$.
    }
    \label{fig:noiseless_tradeoff}
\end{figure}

First, Figure~\ref{fig:noiseless_tradeoff} shows the tradeoff curves of the multi-task system under attenuation factor. For each attenuation level, varying the allocation parameter
$\eta$ generates a task-budget trajectory in the $(L_1,L_2)$ plane.
As the available mean excitation decreases, the curves contract
and become progressively smoother, with the local slope varying gradually
over the allocation range. $\det(A)$ decreases, indicating a
progressive loss of multi-task information capacity. In
contrast, the accompanying variation of $\kappa(A)$ is relatively
moderate and does not produce a strong concentration into a single
collective task mode. Therefore, attenuation primarily reduces the
overall amount of information available for allocation, rather than
fundamentally reshaping its task-directional structure.

\begin{figure}[h]
    \centering
    \includegraphics[width=\linewidth]{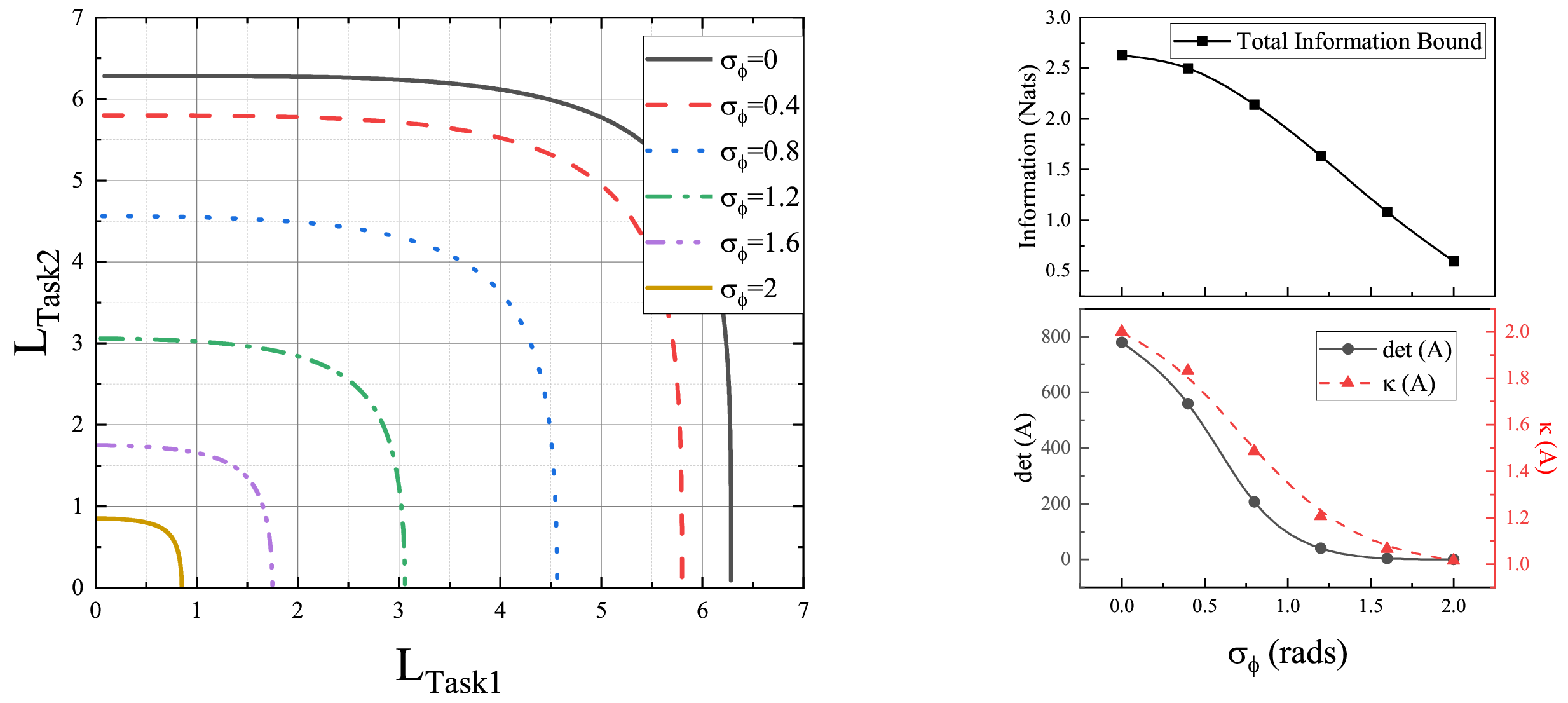}
    \caption{
    Tradeoff curve family under phase diffusion. The curve corresponds to the diffusion factor $\sigma_\phi$ gradually increasing from $0$ to $2$ rads at $N=1/2$. Left: the tradeoff curve family; Right top: the total bound derived by $\mathrm{Tr}A$ at $\eta=1/2$; Right bottom: $\det(A)$ and $\kappa(A)$ at $\eta=1/2$.
    }
    \label{fig:phase_diffusion}
\end{figure}

Second, under phase diffusion, increasing $\sigma_\phi$ suppresses phase coherence and progressively
contracts the tradeoff curves, as can be seen in Figure~\ref{fig:phase_diffusion}, which fixes $N=0.5$ and varies the phase
diffusion strength $\sigma_\phi$. In this case, the capacity decreases, while the allocability remains, as in the attenuation case. This decrease pushes the tradeoff curves inward while retaining flatter edges and a
sharper bending, in contrast to the smoother trend in the attenuation case.

%The observed deformation has two meanings.  First, phase diffusion reduces the total local distinguishability available to the two phase tasks, which appears as a decrease in $T$ and in $B_{\rm tot}(T)$. Second, the spectrum diagnoses whether the degradation is balanced or anisotropic.  In the symmetric diffusion case, the tradeoff mainly contracts while retaining a relatively symmetric form; in asymmetric variants, the same framework would reveal task-selective degradation.

\begin{figure}[h]
    \centering
    \includegraphics[width=\linewidth]{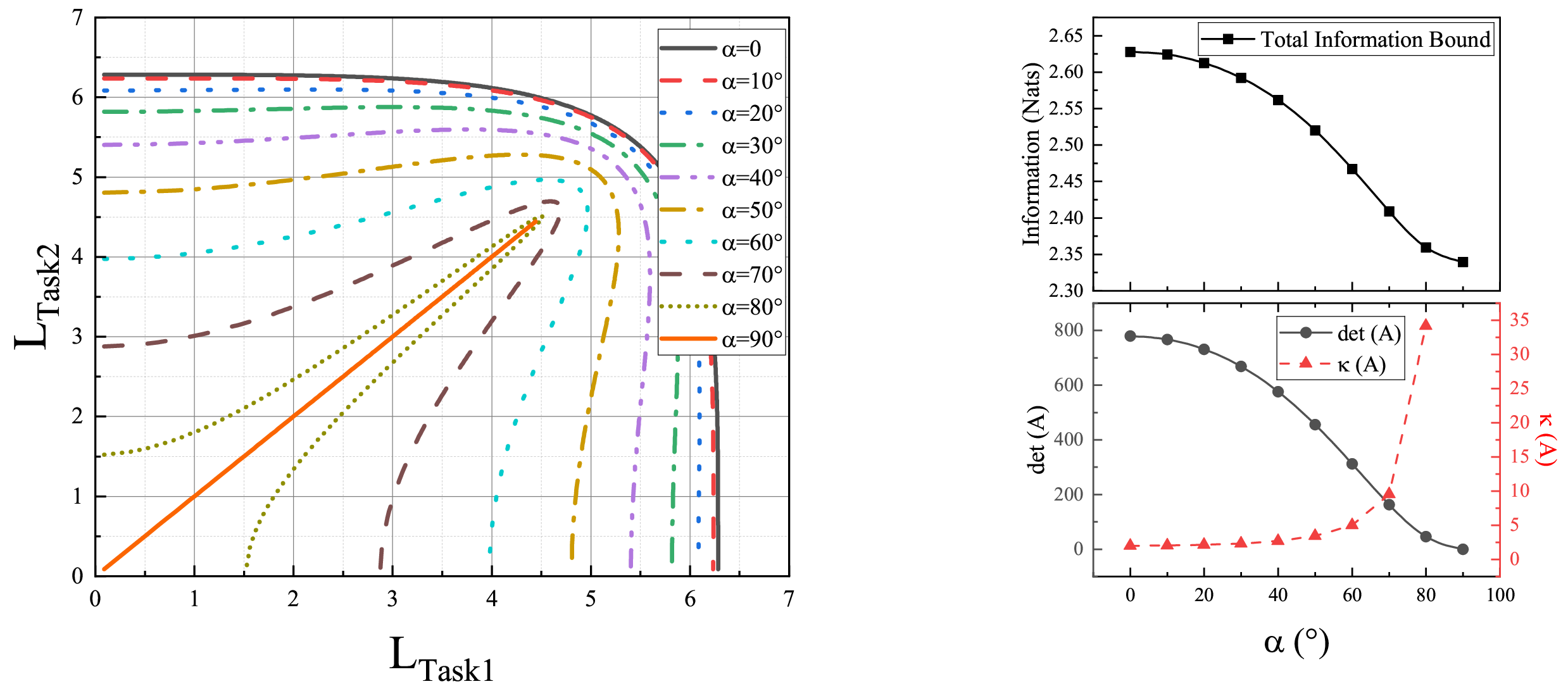}
    \caption{
    Tradeoff curve family under mode crosstalk. The curve corresponds to the mode mixing angle $\alpha$ gradually increasing from $0$ to $90^\circ$ at $N=1/2$. Left: the tradeoff curve family; Right top: the total bound derived by $\mathrm{Tr}A$ at $\eta=1/2$; Right bottom: $\det(A)$ and $\kappa(A)$ at $\eta=1/2$.
    }
    \label{fig:mode_crosstalk}
\end{figure}

Third, we consider the crosstalk between two polarization modes. The decrease in independent mode resource, or the increase of the mode-mixing angle $\alpha$, leads to little capacity decrease near $\eta=1/2$ and much more decrease when $\eta\neq1/2$. Eventually, when $\alpha=90^\circ$, the tradeoff curve shrinks into a line segment. This indicates that crosstalk mainly causes the decrease in allocability and the structural transition, as can be seen in Figure~\ref{fig:mode_crosstalk}. The left
panel shows that crosstalk greatly changes the shape of the tradeoff curves and the right panel shows that the remaining information has become increasingly anisotropic and collective by increasing $\kappa(A)$, while obtaining less independent information by decreasing $\mathrm{det}(A)$.

These non-ideal cases demonstrate that the task tradeoff is not only a property of how the system distributes the physical resource alone. It is jointly induced by the physical carrier, the noise, and the task coordinate direction through the noisy QFIM:
\begin{equation}
\begin{aligned}
    \hat{\rho}_{\text{carrier}}+\mathcal{E}_{\text{channel}}
&\longrightarrow
\mathcal J_Q^{\rm noisy}
\longrightarrow
A\\
&\longrightarrow
\{T,\operatorname{spec}(A),L_1,L_2\}.
\end{aligned}
\end{equation}
Thus loss of performance may appear as a reduction in information amount, a change in information shape, or a combination of both.

%Overall, the simulations support the following interpretation. The total trace-based quantity $T$ controls a coarse joint information ceiling, whereas the task-directional lengths $L_k$ expose how this information is projected onto the prescribed task axes.  The spectrum of $A$ provides an intermediate structural diagnostic: it distinguishes total loss from redistribution or anisotropy of the available information.  Consequently, a task-budget tradeoff is not determined solely by the physical allocation parameter.  Its shape is a property of the full noisy quantum statistical model.

\textit{Conclusion}---
In this work, we have developed a unified, non-asymptotic, measurement-independent, analytically calculable framework for quantum multitask systems, and found an underlying geometric-induced matrix, g-QFIM. It not only benefits the engineering application, but also the theoretical understanding.

% but also reveals the profound geometric nature in information theory.

%The central message is that a multitask tradeoff is not determined by the resource-allocation variable alone: it is jointly shaped by the quantum carrier, the prior geometry, the task coordinates, and the physical noise channel through the effective matrix
%\[
%A=D\mathcal J_QD .
%\]
%Within this framework, \(\operatorname{tr}A\) controls a coarse total information ceiling, \(\operatorname{spec}(A)\) diagnoses the internal structure of the available information, and the diagonal Fisher lengths \(L_k=\sqrt{A_{kk}}\) quantify how this information is exposed along prescribed task directions.

At the application level, this framework provides a way to analyze how different tasks compete under constrained quantum resources in the future quantum network.  It may also be useful to design sensing-assisted communication or privacy-aware multi-task systems, where some parameter directions are intentionally made visible while others are hidden.

At the theoretical level, the present result suggests a profound geometric view of information capacity due to its similarity to Shannon capacity.%, where the channel law and resource constraint determine a maximal information-transfer rate. %Here, however, the relevant local object is not a noise variance or transition probability, but a Fisher metric over the task manifold.
A natural future direction is therefore to formulate tighter bounds as geometric extremal problems. Such a perspective may eventually connect sensing, communication, and geometry in the same way that Shannon theory connects coding, probability, and entropy.

To better support the benefits above, the tightness and achievability of the limit need to be discussed and improved. Additionally, the gap or precise relation between engineering metric and information budget still need further discussion. Overall, this work provides an important step toward a unified fundamental theory for future multi-task application, and is expected to pave the way for future quantum network.

%However, several limitations still remain. First, the established upper bounds are nearly not achievable, and their tightness remains open problem. Additionally, the precise relation between engineering metric and information budget is still not well established. 

%the task-directional curves plotted in this work should be interpreted as allocation-induced information-budget boundaries, not as achievable Pareto frontiers produced by explicit transmitters and receivers.
%Third, our main presentation focuses on rectangular product priors and low-dimensional task structures.  Extending the theory to correlated or irregular priors, as well as to nuisance-aware effective Fisher information and adaptive protocols, is an important next step. 

%Establishing achievability requires specifying concrete measurement, coding, and estimation strategies. 

%\newpage 
\textit{Acknowledgements}---This work was supported by the National Natural Science Foundation of China (62201012, 62571006).

\textit{Data availability}---The data that support the findings of
this Letter are not publicly available. The data are available
from the authors upon reasonable request.

\textit{Conflict of interest statement}---None declared.

%\onecolumngrid

\bibliographystyle{apsrev4-2}
\bibliography{nsr_sample}

\end{document}